\documentclass[fleqn,usenatbib]{mnras}

\usepackage{newtxtext,newtxmath}

\usepackage[T1]{fontenc}

\DeclareRobustCommand{\VAN}[3]{#2}
\let\VANthebibliography\thebibliography
\def\thebibliography{\DeclareRobustCommand{\VAN}[3]{##3}\VANthebibliography}

\usepackage{graphicx}	
\usepackage{amsmath}	

\title[$\nu_{c}$-$r_{cor}$ correlation in Swift J1727.8]{Testing the Lense-Thirring Precession Origin of the QPO in Swift J1727.8--1613}

\author[Ma et al.]{
Ruican Ma,$^{1}$\thanks{E-mail:R.Ma@soton.ac.uk}
Chris Done,$^{2}$
Aya Kubota$^{3}$
\\
$^{1}$School of Physics and Astronomy, University of Southampton, Highfield, Southampton, SO17 1BJ, UK\\
$^{2}$Department of Physics, University of Durham, South Road, Durham, DH1 3LE,UK\\
$^{3}$Department of Electronic Information Systems, Shibaura Institute of Technology, 307 Fukasaku, Minuma-ku, Saitama-shi, Saitama 337-8570, Japan
}

\date{Accepted XXX. Received YYY; in original form ZZZ}

\pubyear{\the\year{}}

\begin{document}
\label{firstpage}
\pagerange{\pageref{firstpage}--\pageref{lastpage}}
\maketitle

\begin{abstract}

We present a comprehensive spectral and timing analysis of the newly discovered black hole transient Swift J1727.8--1613, based on broadband (2--150 keV) observations from \textit{Insight}-HXMT during its 2023 outburst. We use the 
flexible, energy-conserving SS{\sc sed} model
to model both the outer disc and 
inner, complex Comptonisation, using the expected disc emissivity to constrain the corona radius, $r_{cor}$. This decreases from 45\,$R_{\rm g}$ to 9\,$R_{\rm g}$ duing the transition from the hard to hard intermediate and then soft intermediate state.
We plot $r_{cor}$ versus the centroid frequency of the strong quasi-periodic oscillations (QPOs; $\nu_{\rm c}$) seen in these data to test the inner hot flow Lense-Thirring (LT) precession model. The overall slope of the observed trend is in strong agreement with the predictions of LT precession, despite the complexities of accretion behavior, though there is an offset in absolute value which may indicate that the system parameters are still not well determined. The inner radius of the hot flow is consistent with a constant value throughout most of the outburst, indicating that changes in the jet (e.g. the discrete ejections) do not strongly affect the radiated power. Either the jet 
kinetic power is not a large fraction of the accretion power or the jet is instead mostly powered by the spin energy of the black hole.

\end{abstract}




\section{Introduction}

Black hole transients (BHTs) undergo outbursts after months or years of quiescence, lasting from weeks to months \citep[e.g.,][]{Tanaka1996}. These outbursts are marked by transitions between distinct spectral states \citep[for a review, see][]{Homan2005, Belloni2010}, driven by changes in the accretion disc or corona. In the soft state (SS), the multi-temperature disc component dominates, extending down to the innermost stable circular orbit \citep[ISCO;][]{Esin1997}. In contrast, during the hard state (HS), the emission is dominated by Comptonized radiation, with the accretion disc thought to be truncated at tens to hundreds of gravitational radii \citep[$r_{\rm g}$;][]{Done2010}. However, some reports from the reflection modeling suggest that the disc may extend closer to the black hole, with state transitions primarily driven by the changes of the corona \citep{Miller2006, Kara2019}. The intermediate states (IMS) are transitional phases characterized by complex energy spectra, resulting from the significant contributions of both the disc and Comptonization components.

The Comptonization component exhibits spectral characteristics and complex structure across different states. In the HS, it is characterized by a hard photon index ($\Gamma < 1.7$) and a thermal cutoff ranging from tens to hundreds of keV \citep[e.g.,][]{McClintock2006}, whereas in the SS, it shows a much softer Comptonization component ($\Gamma > 2$) with a non-thermal tail extending to higher energies \citep[e.g.,][]{Homan2005}. The IMS exhibits a hybrid Comptonization profile with intricate spectral variations (e.g., Swift J1727.8--1613, \citealt[][]{Liu2024}; XTE J1550--564, \citealt[][]{Kubota2024}; MAXI J1820+070, \citealt[][]{Ma2023}). One potential configuration proposes that both thermal and non-thermal electrons could coexist in the same region \citep{Kubota2024}. Alternatively, thermal electrons reside close to the black hole, while non-thermal electrons are located farther out, potentially accelerated by magnetic flares above the disc \citep{Hjalmarsdotter2016}. 

Fast-time variability provides an independent probe of the accretion process, particularly through quasi-periodic oscillations (QPOs). Strong QPOs, associated with the hard Comptonization component, are typically observed in the bright HS and IMS but weaken in the SS \citep[for a review, see][]{Ingram2019}. Various models have been proposed to explain QPOs, including accretion flow instabilities \citep{Bellavita2022}, jet precession \citep{Ma2021}, and Lense-Thirring precession of the hot flow \citep[hereafter LT model][]{Ingram2009}. Among these, the LT model suggests that QPOs originate from the precession of the whole inner hot flow within a truncated disc \citep{Ingram2009}. This model can be tested by examining the correlation between the truncation radius and QPO frequencies. However, accurately determining the truncation radius remains challenging, particularly given the complex spectral properties in the IMS.

The SS{\sc sed} model is a newly developed, energy-conserving framework that incorporates additional constraints to determine the corona radius, $r_{cor}$, making it especially well-suited for complex energy states \citep{Kubota2024}. It describes a radially stratified accretion flow, 
where there is a standard accretion disc from $r_{out}$ down to $r_{cor}$ which emits a color-temperature corrected blackbody, with the color correction factor ($f_{cor}$) described as in \citet{Done2012}.  Inside $r_{cor}$, the accretion power is dissipated via inverse Compton scattering, where seed photons originate from reprocessing in a passive disc underlying the corona. Observations show that this Comptonised emission is often more complex than can be described by a single temperature thermal Compton component, so the SS{\sc sed} model includes two thermal electron distributions, characterized by the parameters $kTe_{1,2}$ and $\Gamma_{1,2}$, respectively, though the highest temperature component is more likely physically to indicate a non-thermal tail to the electron distribution. The relative luminosity in each of these two components is parameterised by $f_{th}$ which is the fraction of lowest temperature 
Comptonisation to total Comptonization. The model ensures energy conservation by constraining the radiation to follow the standard accretion disc emissivity \citep{Shakura1973}, assuming a constant mass accretion rate ($\dot{M}$) across all radii (from the outer disc radius, $r_{\rm out}$, to the inner disc radius, $r_{\rm in}$). In addition, the SS{\sc sed} model requires key system parameters of mass ($M$), distance ($D$), and inclination ($i$) to estimate the energy spectra shape and luminosity. \cite{Kubota2024} analyzed RXTE data of XTE~J$1550-564$, and compared the estimated values of $r_{cor}$ to their QPO centroid frequencies $\nu_{\rm c}$. They found that the derived $\nu_{\rm c}$-$r_{cor}$ relation is well consistent with the LT precession scenario by \cite{Ingram2009} by interpreted $r_{cor}$ as inner hot-torus radius, i.e., the disc truncation radius $r_{tr}$.

Swift J1727.8--1613 was identified as a new Galactic transient on August 23, 2023, as its flux rapidly increased, reaching up to 7.6 Crab in the 15--50 keV \citep{Palmer2023}. Subsequent observations indicate that this source exhibits characteristics of a BHT, including X-ray \citep{Nakajima2023, Draghis2023, Peng2024}, optical \citep{Castro-Tirado2023, Wang2023}, and radio \citep{Miller-Jones2023a, Miller-Jones2023b} emissions. \citet{Mata2024b}  reported the distance as $3.4 \pm 0.3$ kpc based on various empirical methods using the 10.4-m Gran Telescopio Canarias optical data. Swift J1727.8--1613 may possess a high inclination, as inferred from the properties of its type-C QPOs, analyzed with \textit{Insight}-HXMT data \citep{Yu2024,Zhu2024}, and polarization analysis \citep{Svoboda2024a,Svoboda2024b}. The source exhibits strong type-C QPOs up to 150 keV \citep{Yu2024, Zhu2024, Yang2024}, and several studies have further explored their characteristics in connection with the properties of the corona \citep[e.g.,][]{Rawat2025, Liao2025}. Additionally, the source shows intriguing X-ray characteristics: Swift J1727.8--1613 undergoes significant flaring during the outburst, referred to as the flare state \citep{Yu2024, Liu2024}. In addition to the disc and primary Comptonization components, Swift J1727.8–1613 exhibits more complex X-ray spectra, featuring an additional hard tail. This component, well described by a Comptonization model, has been detected by \textit{INTEGRAL}, \textit{NuSTAR}, and \textit{Insight}-HXMT \citep{Mereminskiy2024, Peng2024, Liu2024}.
Polarization results indicate that the corona of the source extends in the disc plane, as analyzed from \textit{IXPE} data during HS and transitions from HS to SS \citep{Veledina2023, Ingram2024}. Additionally, \citet{Zhao2024} conducted the first polarimetric analysis of QPOs in this source using \textit{IXPE} data, revealing significant polarization degrees in the IMS that are independent of the QPO phase.

In this work, we employ the SS{\sc sed} model to derive the corona radius, $r_{cor}$, and use it to solely identify $r_{tr}$ in the complicated spectrum in the IMS. We then investigate the theoretical relation between $r_{cor}$ and QPO centroid frequency ($\nu_{\rm c}$) to test the LT model. Our analysis is based on nearly 300 broadband observations (2--150\,keV) of the newly discovered black hole candidate (BHC), Swift J1727.8--1613, conducted by \textit{Insight}-HXMT. The paper is structured as follows: Section~\ref{sec:data} describes the observations used and the data reduction, Section~\ref{sec:results} presents the energy spectral fitting results using the SS{\sc sed} model, timing results and the relation between truncated radius and QPO centroid frequency, Section~\ref{sec:dis} discusses our results and evaluates the validity of SS{\sc sed} model and LT model, and Section~\ref{sec:sum} provides a summary.

\section{Observation and data reduction}
\label{sec:data}

\textit{Insight}-HXMT (hereafter HXMT) is China's first X-ray astronomy satellite, launched in June 2017, and it has been in orbit for more than 7 years \citep{Zhang2020}. HXMT carries three instruments: the Low-Energy \citep[LE, 1--15\,keV, 384\,${\rm cm^{2}}$;][]{Chen2020}, Medium-Energy \citep[ME, 5--30\,keV, 952\,${\rm cm^{2}}$;][]{Cao2020}, and High-Energy \citep[HE, 20--250\,keV, 5100\,${\rm cm^{2}}$;][]{Liu2020} X-ray telescopes. In addition to its broadband energy range (1--250\,keV) and large effective area, HXMT is particularly suitable for observing bright sources due to its no pileup effect.

Following observations by \textit{MAXI} and \textit{Swift}, HXMT monitored Swift J1727.8--1613 from August 25, 2023, to October 6, 2023 (MJD 60181 to MJD 60223), corresponding to ObsIDs P0614338001--P0614338035. Then, due to solar obscuration, the source could not be observed by HXMT again until March 1, 2024 (MJD 60370). However, because of the low flux of the source, only one additional observation, ObsID P0614338040, was performed by HXMT before the source finished its outburst. In this study, we present results exclusively for periods exhibiting strong QPOs, specifically utilizing HXMT data from MJD 60181 to MJD 60223.

We reduced the data using the {\it hpipeline} of the HXMT Data Analysis software ({\sc hxmtdas}) version 2.06. The data were filtered according to the recommendations of the HXMT team: (1) Earth elevation angle >$10^{\circ}$; (2) geomagnetic cutoff rigidity >8\,GV; (3) at least 300\,s before and after passage through the South Atlantic Anomaly; (4) pointing offset angle <$0.04^{\circ}$. To avoid possible contamination from the bright Earth and nearby sources, we used data only from the small field-of-view detectors. The backgrounds for the LE, ME, and HE were estimated using the tools LEBKGMAP, MEBKGMAP, and HEBKGMAP, respectively, based on the HXMT background model \citep{Guo2020, Liao2020a, Liao2020b}.

\begin{figure*}
\centering
\includegraphics[width=0.9\textwidth]{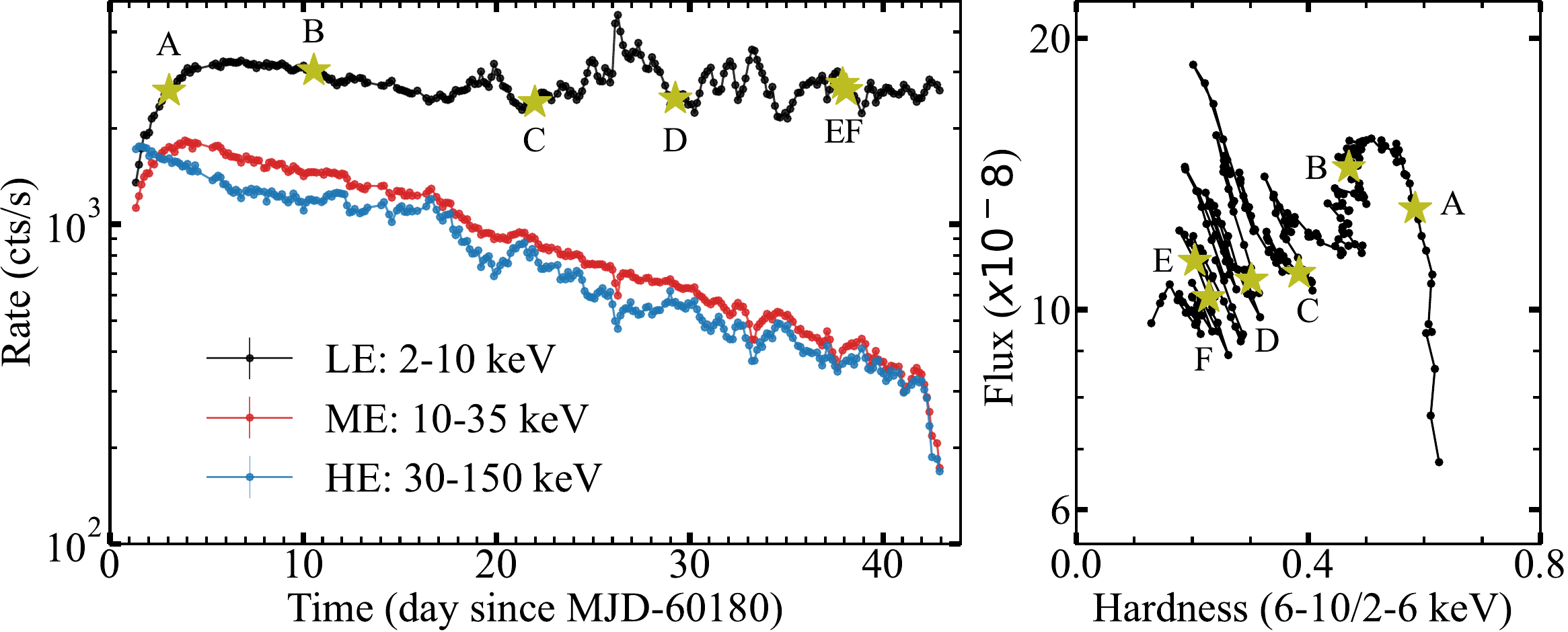} \\
\caption{Left panel: HXMT light curve of Swift J1727.8–1613, with black, red, and blue points representing the LE (2--10\,keV), ME (10--35\,keV), and HE (30--150\,keV) bands, respectively. Right panel: Hardness-intensity diagram (HID) of the source, where hardness is defined as the count rate ratio between the 6--10\,keV and 2--6\,keV, while intensity corresponds to the unabsorbed flux in the 2--10\,keV. The yellow star marks the representative ObsIDs A--F \textsuperscript{1} illustrated in Fig.~\ref{fig:pha}. }
\label{fig:lc_hid}
\end{figure*}
\footnotetext[1]{The ObsIDs corresponding to A through F are: P061433800201, P061433800512, P061433801507, P061433802302, P061433803107, and P061433803201, respectively.}

\begin{figure*}
\includegraphics[width=\textwidth]{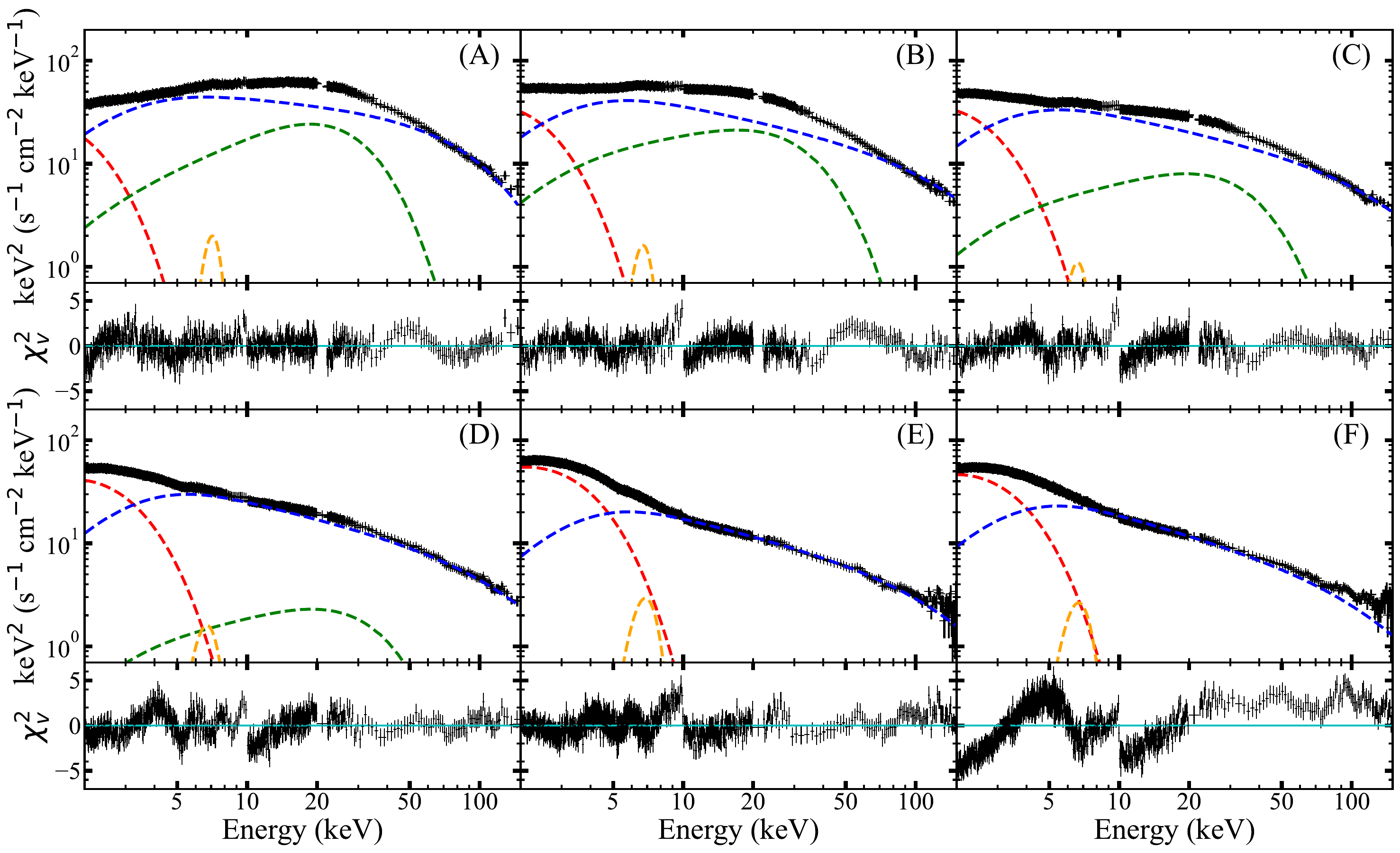} \\
\caption{Representative best-fit energy spectra. Panels (a) to (f) correspond to different spectral phases: HS, HIMS1, FS1, FS2, HIMS2, and HIMS3, respectively. The red, blue, green, and orange dashed lines represent the disc component, non-thermal/thermal Comptonization components, and the Gaussian line, respectively.
\label{fig:pha}
}
\end{figure*}

\begin{figure*}
\includegraphics[width=\textwidth]{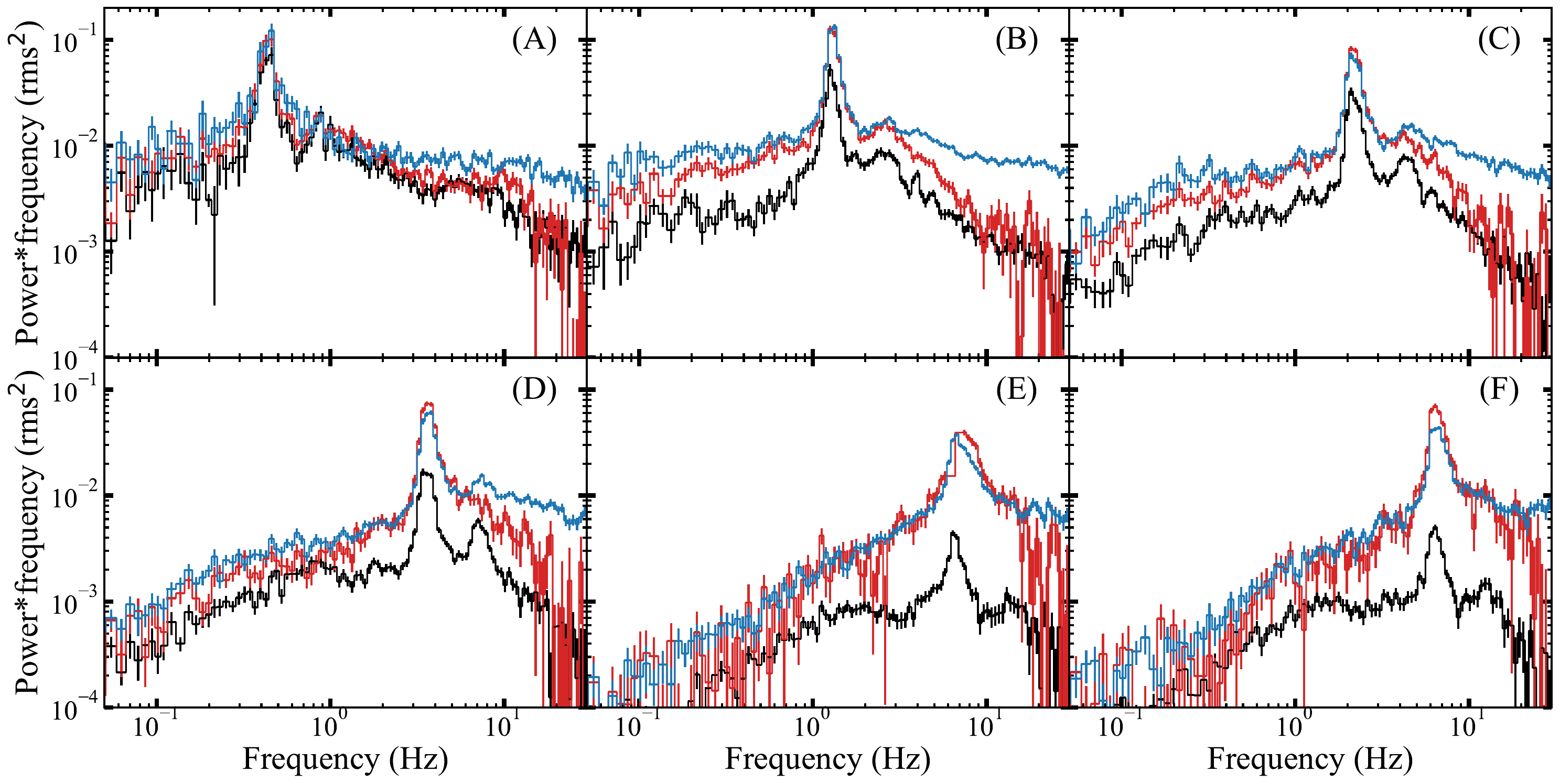} \\
\caption{PDS corresponding to Fig.~\ref{fig:pha}, panels (a)–(f), which represent different spectral states: HS, HIMS1, FS1, FS2, HIMS2, and HIMS3, respectively. In each panel, the black, red, and blue denote data from the LE, ME, and HE, respectively
\label{fig:pds}
}
\end{figure*}

\begin{table*}
\renewcommand{\arraystretch}{1.3}
\label{tab:pars}
\begin{tabular}{llllllll}
\hline
Component & Parameter & Obs. A  &  Obs. B & Obs. C  & Obs. D & Obs. E & Obs. F \\ \hline
{\tt SS{\sc sed}}  &  log $\dot{m}$ (${\dot Mc^{2}}/{L_{\rm Edd}}$)  &  $0.5957\pm{0.0007}$ &  $0.6005\pm{0.0006}$ &  $0.4809\pm{0.0005}$ & $0.4445\pm{0.0005}$ & $0.4261\pm{0.0004}$ & $0.3947\pm{0.0005}$ \\ 
  &  $r_{\rm in}$ ($r_{\rm g}$)  &  $[4.5]^{\dag}$ &  $[4.5]^{\dag}$ &  $[4.5]^{\dag}$ & $[4.5]^{\dag}$ & $[4.5]^{\dag}$ & $[4.5]^{\dag}$ \\  
  &  $R_{cor}$ ($r_{\rm g}$)  &  $42.3\pm{0.3}$ &  $30.19\pm{0.16}$ &  $23.32\pm{0.08}$ & $17.46_{-0.07}^{+0.06}$ & $11.50\pm{0.02}$ & $13.05_{-0.04}^{+0.03}$ \\ 
  &  $kT_{\rm e1}$ (keV)  &  $5.68_{-0.05}^{+0.08}$ &  $6.8\pm{0.2}$ &  $7.19\pm{0.11}$ & $[7]^{\dag}$ & - & - \\
  &  $kT_{\rm e2}$ (keV)  &  $33.7_{-1.9}^{+2.2}$ &  $[300]^{\dag}$ &  $[300]^{\dag}$ & $[300]^{\dag}$ & $[300]^{\dag}$ & $[300]^{\dag}$ \\ 
  &  $\Gamma_1$  &  $1.40_{-0.00*}^{+0.02}$ &  $1.67\pm{0.05}$ &  $[1.6]^{\dag}$ & $[1.6]^{\dag}$ & - & -\\ 
  &  $\Gamma_2$  &  $2.175\pm{0.010}$ &  $2.4161_{-0.010}^{+0.009}$ &  $2.434\pm{0.004}$ & $2.492\pm{0.003}$ & $2.588\pm{0.004}$ & $2.508\pm{0.005}$ \\ 
  &  $f_{th}$  &  $0.237_{-0.006}^{+0.010}$ &  $0.291_{-0.020}^{+0.021}$ &  $0.158\pm{0.003}$ & $0.059\pm{0.004}$ & - & -\\ \hline
{\tt Gaussian}  &  $E_c$ (keV)  &  $7.00\pm{0.16}$ &  $6.63_{-0.14}^{+0.16}$ &  $6.54\pm{0.15}$ & $6.55\pm{0.08}$ & $6.69\pm{0.05}$ & $6.500_{-0.000*}^{+0.002}$ \\  
  &  $\sigma$ (keV)  &  $0.55_{-0.16}^{+0.17}$ &  $0.56_{-0.14}^{+0.16}$ &  $0.56\pm{0.12}$ & $0.72_{-0.10}^{+0.08}$ & $0.800_{-0.004}^{+0.000*}$ & $0.800_{-0.003}^{+0.000*}$ \\ 
  &  $norm$ ($\times 10^{-2}$)  &  $5.5_{-1.4}^{+1.5}$ &  $5.1_{-1.2}^{+1.3}$ &  $3.6\pm{0.8}$ & $6.6_{-1.0}^{+1.1}$ & $12.8\pm{0.5}$ & $12.1_{-0.6}^{+0.7}$ \\ \hline 
  &  $\chi^2/dof$  &  1399/1404 &  1293/1405 &  1541/1406 & 1571/1405 & 1541/1406 & 5289/1406 \\ \hline 
  &  $\nu_{\rm c}$ (Hz)  &  $0.431_{-0.013}^{+0.012}$ &  $1.265\pm{0.011}$ &  $2.203\pm{0.017}$ & $3.604_{-0.022}^{+0.013}$ & $6.53_{-0.05}^{+0.06}$ & $6.356_{-0.044}^{+0.006}$ \\ \hline

\end{tabular}
\caption{Spectral parameters of the representative best-fitting spectra for Observations A--F. The values of hydrogen column density ($N_{\rm H}$), black hole mass, distance, and inclination angle are fixed at $2\times 10^{21}\,{\rm cm}^{-2}$, 9\,$M_{\odot}$, 3.4\,kpc, and $60^{\circ}$, respectively.  
$^\dag$: Parameter value is fixed due to poor constraints. 
$^*$:  The positive or negative error is pegged at the upper or lower limit.}
\end{table*}

\begin{figure*}
\centering
\includegraphics[width=0.95\textwidth]{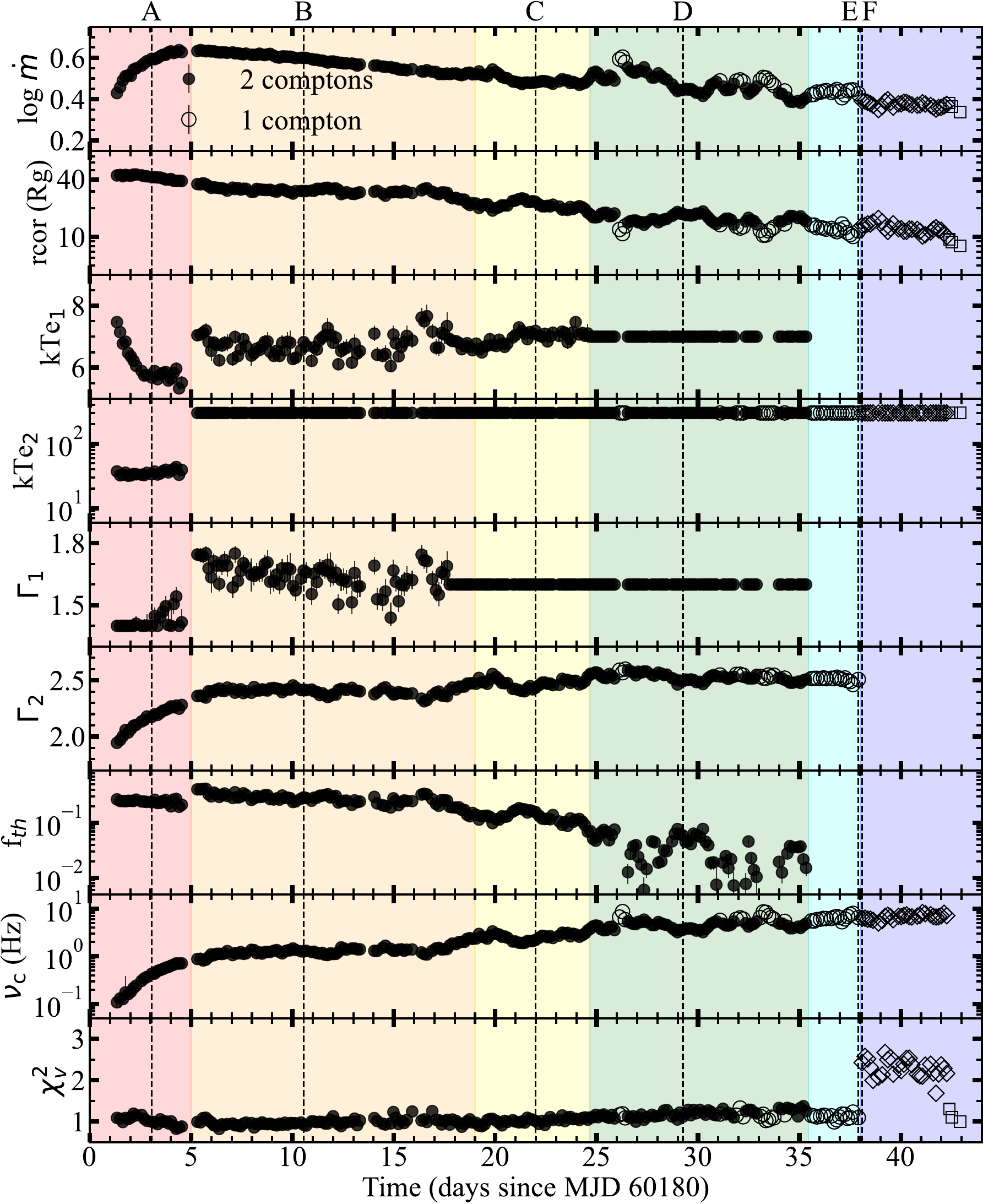} \\
\caption{The evolution of the best-fitting spectral parameters for Swift J1727.8--1613. From top to bottom, the panels display: the logarithm of the accretion rate (log\,$\dot{m}$, where $\dot{m}={\dot Mc^{2}}/{L_{\rm Edd}}$), the corona radius ($r_{cor}$) in $R_{\rm g}$. 
$kT_{\rm e1}$ and $kT_{\rm e2}$ are the electron temperature of the thermal and non-thermal\textsuperscript{2} Comptonization components, respectively, $\Gamma_{1}$ and $\Gamma_{2}$ are the corresponding photon indices.
The fraction of the thermal Comptonization component relative to the total Comptonization ($f_{th}$), the QPO centroid frequency ($\nu_{\rm c}$), and the reduced chi-squared values ($\chi^{2}_{v}$), respectively. 
Filled symbols indicate observations that require two Comptonization components for a good fit, while open symbols correspond to cases where only the non-thermal Comptonization component is needed. Circles represent data with QPOs, and squares represent data without QPOs.
The red, orange, yellow, green, cyan, and blue shaded regions represent HS, HIMS1, FS1, FS2, HIMS2, and HIMS3, respectively. Letters A--F correspond to the representative ObsIDs shown in Fig.~\ref{fig:pha}. 
\label{fig:pars}
}
\end{figure*}
\footnotetext[2]{In the hard state, the emission is dominated by thermal, rather than non-thermal, Comptonization.}

\section{Results}
\label{sec:results}

\subsection{Overview of the outburst}
\label{subsec:outburst}

We present the HXMT light curve and hardness-intensity diagram (HID) of Swift J1727.8--1613 during its 2023 outburst in Fig.~\ref{fig:lc_hid}. The LE (2--10\,keV) count rate increased rapidly from approximately 1400\,cts\,s$^{-1}$ on MJD 60181 to around 3200\,cts\,s$^{-1}$ on MJD 60186. It then slowly decreased to about 2400\,cts\,s$^{-1}$ on MJD 60196, followed by multiple flares, referred to as the 'flare state' \citep{Yu2024,Liu2024}, during which the count rate remained relatively stable. The ME count rate (10--35\,keV) increased from approximately 1100\,cts\,s$^{-1}$ on MJD 60180 to about 1800\,cts\,s$^{-1}$ on MJD 60184, then showed a slow decline, decreasing to around 180\,cts\,s$^{-1}$ on MJD 60223. The HE count rate (30--150\,keV) showed a declining trend throughout the HXMT observation period, from approximately 1800\,cts\,s$^{-1}$ on MJD 60181 to about 170\,cts\,s$^{-1}$ on MJD 60223. The total luminosity changes by only a factor of 2 across all these datasets. 

The HID is presented in the right panel of Fig.~\ref{fig:lc_hid}. The hardness is defined as the ratio of the LE count rate in the 6--10\,keV energy band to the LE count rate in the 2--6\,keV energy band, and the intensity is the unabsorbed flux in the 2--10\,keV. At the beginning of the HXMT monitoring, the LE flux increased rapidly by a factor 2, while the hardness remained constantly high at around 0.6. This indicates that the source was in the HS, as reported by \citet{Liu2024}. Subsequently, the hardness decreased from approximately 0.6 to 0.1, with a flaring flux, indicating that the source transitioned to the IMS. Moreover, the hardness always remained higher than 0.03, which implies that Swift J1727.8--1613 has a high inclination \citep{Munoz2013}, which is consistent with the reported by \citet{Yu2024} and \citet{Liu2024}.

We select 6 representative spectra across the outburst as examples of the source behaviour, marked A--F on Fig.~\ref{fig:lc_hid}, Fig.~\ref{fig:pha} shows their energy spectra (see below), while Fig.~\ref{fig:pds} shows their power density spectra (PDS). Clearly there is considerable evolution of the source spectrum and timing properties throughout this period.

\subsection{Energy spectral fitting}

We analyzed the energy spectra using {\sc xspec} v12.13.1 \citep{Arnaud1996}. The energy bands utilized in this study were defined as follows: 2--10 keV for the LE, 10--20 keV, and 22--35 keV for the ME \citep[excluding the 20--22 keV energy band due to contamination from the silver fluorescence line;][]{Guo2020}, and 30--150 keV for the HE. Systematic errors of 1\%, 1\%, and 2\% were applied to the LE, ME, and HE instruments, respectively \citep{Liao2020a, Guo2020, Liao2020b}.

We use the {\tt tbabs} model for neutral absorption from the interstellar medium in the direction of the source, employing the abundance and cross-section tables from \citet{Wilms2000} and \citet{Verner1996}, respectively. We fix the hydrogen column density along the line of sight to the source, $N_{\rm H}$, at $2 \times 10^{21}\,{\rm cm^{-2}}$, based on the average value reported by \citet{HI4PI2016}.
We use the energy-conserving model {SS\sc{sed}} to describe the emission from the disc and inhomogeneous Comptonization components. Considering the strong iron emission line of Swift J1727.8--1613, we added a {\tt Gaussian} to the energy spectral fitting. The total model is {\tt tbabs*(SS$_{\rm SED}$ + Gaussian)} in {\sc xspec}. 

We assume a mass of 9\,$M_{\rm \odot}$ and a inclination, $i$, of $60^{\circ}$, conseriding that this source probably possess a high inclination \citep[e.g.,][]{Yu2024,Zhu2024}. The distance is set to 3.4\,kpc \citep{Mata2024b}. 
Additionally, we apply a color correction factor, as described in \citet{Done2012} (by setting the code parameter flag  $color\_cor$, to 1).

The remaining parameters in {SS\sc{sed}}, including the accretion rate ($\dot M$), inner disc radius ($r_{\rm in}$), electron temperatures of the thermal ($kT_{\rm e1}$) and non-thermal ($kT_{\rm e2}$) Comptonization components, photon indices of thermal ($\Gamma_{1}$) and non-thermal ($\Gamma_{2}$) Comptonization components, a fraction of the thermal Comptonising component to the total Comptonization ($f_{th}$), and outer radius of the disc-corona $r_{cor}$ (i.e., disc-corona truncated radius ) are freely fitted or fixed according to the physical properties of Swift J1727.8--1613 in different energy states (details are provided in the following subsections).

\subsubsection{The disc dominated state: inner disc radius}

There is only one spectrum which satisfies the requirement to be defined as a disc dominated soft state, and that is on day 42, ObsID=P0614338040.
We fit this with the mass, distance and inclination above, and use a single Comptonisation model to describe the weak tail. This gives $r_{\rm in}=4.5$
so we initially fix this across all the datasets. 

\subsubsection{The hard state}

We start fitting at the start of the outburst in the 
hard state. This spans day 1.3--4.5 (the red shaded area in Fig.~\ref{fig:pars}, ObsID P061433800101--P061433800212), during which the bolometric luminosity increases from $\sim3.7\times 10^{38}$\,erg\,s$^{-1}$ ($\sim 0.32\,L_{\rm Edd}$) to $\sim5.9\times 10^{38}$\,erg\,s$^{-1}$ ($\sim 0.50\,L_{\rm Edd}$). The hard state spectrum can be phenomenologically fitted using the SS{\sc sed} model by two distinct thermal components \citep[similar with XTE J1550--564 in][]{Kubota2024}. However, as noted in \citet{Kubota2024} and \citet{Kubota2018}, reproducing a hard spectrum with photon index $\Gamma \textless 2$ is challenging within the framework of a passive disc plus corona. Since the SS{\sc sed} model assumes a passive disc, the estimation of the coronal radius ($r_{cor}$) is subject to substantial systematic uncertainties. Therefore, the derived values in the hard state should be regarded as reference values only. Moreover, given the weak disc emission in this state, particularly above 2\,keV, we maintain $r_{\rm in}$ fixed at 4.5 for consistency. The representative spectra and PDS are shown in panel (a) of Figs~\ref{fig:pha} and \ref{fig:pds}, with the corresponding spectral parameters listed in Table~\ref{tab:pars}.

In our spectral fitting, the two thermal Comptonization components exhibit distinct properties. The hotter component shows a slight increase in electron temperature, $kT_{\rm e2}$, from $\sim$32\,keV to $\sim$44\,keV, accompanied by a softening photon index ($\Gamma_2$ increasing from $\sim$1.9 to $\sim$2.3). Meanwhile, the cooler component ($kT_{\rm e1}$) remains within 5--8\,keV, with a harder photon index ($\Gamma_1 \sim 1.4-1.7$). 
The $f_{th}$ remains nearly constant at 0.25, while $r_{cor}$ gradually decreases from 45\,$R_{\rm g}$ to 39\,$R_{\rm g}$. 

The shape of the cooler, harder Comptonisation component is similar to the shape expected from low ionisation reflected emission. 
We show specific fits with reflection in the appendix~\ref{sec:app_B}, but we find that the broadband spectral curvature cannot be well fit with just a single thermal Comptonisation component and its reflection. This is also shown in other sources where the spectra span a broad bandpass e.g. \citet{zdziarski_2021}.

\subsubsection{The hardest HIMS }
\label{sec:hims}

Most of the remainder of the outburst from day 5.3 to day 42 is in the HIMS \citep{Bollemeĳer2025}, but here we 
further subdivide the data. We designate
HIMS1 from day 5.3 to MJD 18.9 (the yellow and orange shaded areas in Fig.~\ref{fig:pars}, ObsID P061433800301--P061433801207). During this phase, the luminosity ranges from $\sim6.0\times 10^{38}$\,erg/s ($\sim 0.52\,L_{\rm Edd}$) to $\sim4.6\times 10^{38}$\,erg/s ($\sim 0.40\,L_{\rm Edd}$). In the spectral fitting, the parameters $f_{th}$, $r_{cor}$, and $\dot{M}$ are left free, whereas the electron temperature of the non-thermal Comptonization corona is set to $kT_{\rm e2} = 300$\,keV\footnote{The fixed $kT_{\rm e2}$ value provides a reasonable approximation of the energy spectra, as the break energy of the non-thermal Comptonization component—determined by the Klein–Nishina cross-section \citep{Hjalmarsdotter2016}—is 511\,keV, whereas our observational data extend only up to 150\,keV.}. 
The representative spectra and PDS are shown in panel (b) of Figs~\ref{fig:pha} and ~\ref{fig:pds}, with the corresponding spectral parameters listed in Table~\ref{tab:pars}.

During HIMS1, $r_{cor}$ decreases gradually from 37\,$R_{\rm g}$ to 22\,$R_{\rm g}$, accompanied by a significant reduction in $f_{th}$ from 0.42 to 0.13. The photon index of the non-thermal Comptonization component, $\Gamma_2$, remains relatively constant at 2.4, while the photon index of the thermal Comptonization component, $\Gamma_1$, decreases from 1.8 to 1.4. The electron temperature $kT_{\rm e1}$ of the thermal component remains stable at around 7\,keV. Additionally, the optical depth $\tau_{\rm es}$ and $y$-parameter of the thermal Comptonization vary between 7.5--8.7 and 3.0--4.1, respectively.

\subsubsection{Flare state}

There is substantial flaring from day 19.1--35.3, (ObsID P061433801301--P061433802903)
so this has been called the flaring state \citep{Liu2024,Yu2024}. These data are generally well-fitted with a combination of thermal and non-thermal Comptonization coronae. During this period, the luminosity varies between $\sim3.3\times 10^{38}$\,erg/s ($\sim 0.29\,L_{\rm Edd}$) and $\sim5.0\times 10^{38}$\,erg/s ($\sim 0.43\,L_{\rm Edd}$).  Since the low-energy band is limited to 2 keV, and we also do not expect significant variation in the inner disc radius, $r_{\rm in}$, therefore, we fix $r_{\rm in} = 4.5$ during spectral fitting. In most cases, both Comptonization components are required, while some observations near the peak flux in the LE data can be well-fitted with only a single non-thermal Comptonization component.

The representative spectra and PDS are shown in Figs~\ref{fig:pha} and ~\ref{fig:pds}, panels (c) and (d), with the corresponding spectral parameters listed in Table~\ref{tab:pars}. The evolution of these parameters is presented in Fig~\ref{fig:pars}. The best-fitting model for the flare state is similar to that of the HIMS1, comprising both non-thermal ($kT_{\rm e2} = 300$\,keV) and thermal Comptonization components. The photon index $\Gamma_2$ remains approximately within the range of 2.4 to 2.6. The weak thermal Comptonization component, with a photon index fixed at 1.6, corresponds to the average value observed in the hard intermediate state (see Section~\ref{sec:hims}).

To facilitate analysis, we further divide this period into two phases:  FS1 (ObsID P061433801301--P061433801804, day 19.1--24.6, green) and phase FS2 (ObsID P061433801805--P061433802903, day 24.7--35.3, yellow). During phase FS1, the electron temperature of the weak thermal Comptonization component, $kT_{\rm e1}$, remains stable at approximately 7 keV. However, in phase FS2, this parameter is poorly constrained; therefore, we fix it at 7 keV to ensure the robustness of the spectral fitting.

The fitting results indicate that in phase FS1, $r_{cor}$ gradually decreases from $\sim26\,R_{\rm g}$ to $\sim18\,R_{\rm g}$, accompanied by a decrease in $f_{th}$ from 0.2 to 0.07. In phase FS2, as $r_{cor}$ further shrinks from $\sim18\,R_{\rm g}$ to $\sim13\,R_{\rm g}$, $f_{th}$ remains consistently low, below 0.1.

In addition, several observations\footnote{Total 14 ObsIDs of the observation can be fit with a single non-thermal Comptonization: P061433802002--P061433802004, P061433802501, P061433802507, P061433802601--P061433802602, P061433802701--P061433802706, P061433802803} in phase FS2 exhibit an extremely low $f_{th}$ ($<0.01$), indicating that the thermal Comptonization component is no longer required. As shown in Fig~\ref{fig:pars}, these spectra can be well fitted with a single non-thermal Comptonization model, accompanied by a relatively small $r_{cor}$ and a high accretion rate. Specifically, all of these observations correspond to the peak accretion rate and the softest hardness ratio during the flare state, suggesting that the increase in accretion rate leads to a stronger non-thermal Comptonization component, which suppresses the thermal Comptonization and results in a softer spectrum.

\subsubsection{The softer HIMS: changing inner disc radius}

The final phase of the observation is still in apparently in the HIMS (ObsID P061433802905--P061433803503, day 35.6--42.3. The spectrum is softening but the QPO is still type C and still accompanied by some broad band noise. 
The spectra up to day 38 (blue shaded area in Fig~\ref{fig:pars}, HIMS2) is well fit by a single 
steep non-thermal Comptonization component ($kT_{\rm e2}$ = 300\,keV, $\Gamma_2 \sim 2.5$), without requiring a significant thermal Comptonization contribution. Consequently, we set $f_{th} = 0$. 

However, after day 38  there is a sudden jump in reduced chi-squared from $\approx 1$ to $\ge 2$ until the source approaches the soft state after day 42 (see Fig.~\ref{fig:pars}). We use this to mark HIMS3 (ObsID P061433803201--P061433803503, day 38.6--42.3, purple shaded area in Fig~\ref{fig:pars}).
An abrupt change in source behaviour at this point is also noted by \cite{Stiele2024}. The data show a marked jump in the disc shape, with the peak normalization dropping and the disc appearing slightly broader, as if more Comptonised perhaps by low temperature material. There is very little change in the coronal emission above 5~keV. We find that we can get a good fit in HIMS3 by relaxing the assumption of a constant $r_{\rm in}$ at $4.5$. We refit the entire HIMS2-3 period with a free $r_{\rm in}$ and find that 
the model fits with  an abrupt decrease in $r_{\rm in}$ to $\sim 3.2$ at day 38 . However this is accompanied by an abrupt drop in overall mass accretion rate to compensate for the increased energy available from a flow extending closer to the black hole. We show these fits in the Fig.~\ref{fig:pars_appendix} and \ref{fig:lt_appendix} in Appendix A, but prefer to keep here to a model with fixed $r_{\rm in}$ and smoothly varying mass accretion rate. 

\subsection{Timing results}

To determine the QPO centroid frequency of Swift J1727.8–1613, we used {\sc powspec} in the \textit{XRONOS} software to obtain the PDS in the 1--10 keV energy band. The segment length and time resolution for each Fast Fourier Transform (FFT) were set to 128\,s and 3.9\,ms, respectively, corresponding to a minimum frequency of 0.0078125\,Hz and a Nyquist frequency of 128\,Hz. We subtracted the Poisson noise and normalized the PDS using the Miyamoto normalization \citep{Miyamoto1991}. The PDS was then fitted with Lorentzian functions to model one or more noise components and one or more peaks, following a similar model to \citet{Yu2024}.  

The representative PDS for different spectral states, corresponding to Fig.~\ref{fig:pha}, are presented in Fig.~\ref{fig:pds}. The evolution of the QPO centroid frequency is shown in Fig~\ref{fig:pars}. It gradually increases from 0.1\,Hz to 8.0\,Hz. In some observations with high accretion rates, the QPO centroid frequency rises more abruptly, reaching up to 8.6\,Hz. 

A more detailed study of the QPO properties (including cross-spectra and lags) is given by \citet{Yu2024} and \citet{Bollemeĳer2025}.

\section{Discussion}
\label{sec:dis}
We analyzed the energy spectral and timing properties of Swift J1727.8--1613 using 265 broadband HXMT observations in the 2--150 keV energy band. By applying the new energy-conserving spectral model, SS{\sc sed}, we fit the energy spectrum and determine the radius at which the disc power is dissipated as Comptonisation rather than thermalizing to a local (colour temperature corrected) blackbody. 
This radius, $r_{cor}$, is determined from the data by energy balance: the disc outside of  $r_{cor}$ has luminosity $L_{\sc disc}$ while the complex Comptonisation has luminosity $L_{\sc cor}$, so that 
\begin{equation}
\frac{L_{\rm cor}} {L_{\rm cor}+L_{\rm disc}}= \frac{\int_{r_{cor}}^{r_{\rm in}} \epsilon_{SS}(r)4\pi rdr }{\int_\infty^{r_{\rm in}} \epsilon_{SS}(r)4\pi rdr }
\end{equation}
subject also to the constraint that the disc luminosity and temperature at  $r> r_{cor}$ is consistent with the derived $\dot{M}$.

\begin{figure*}
\centering
\includegraphics[width=\textwidth]{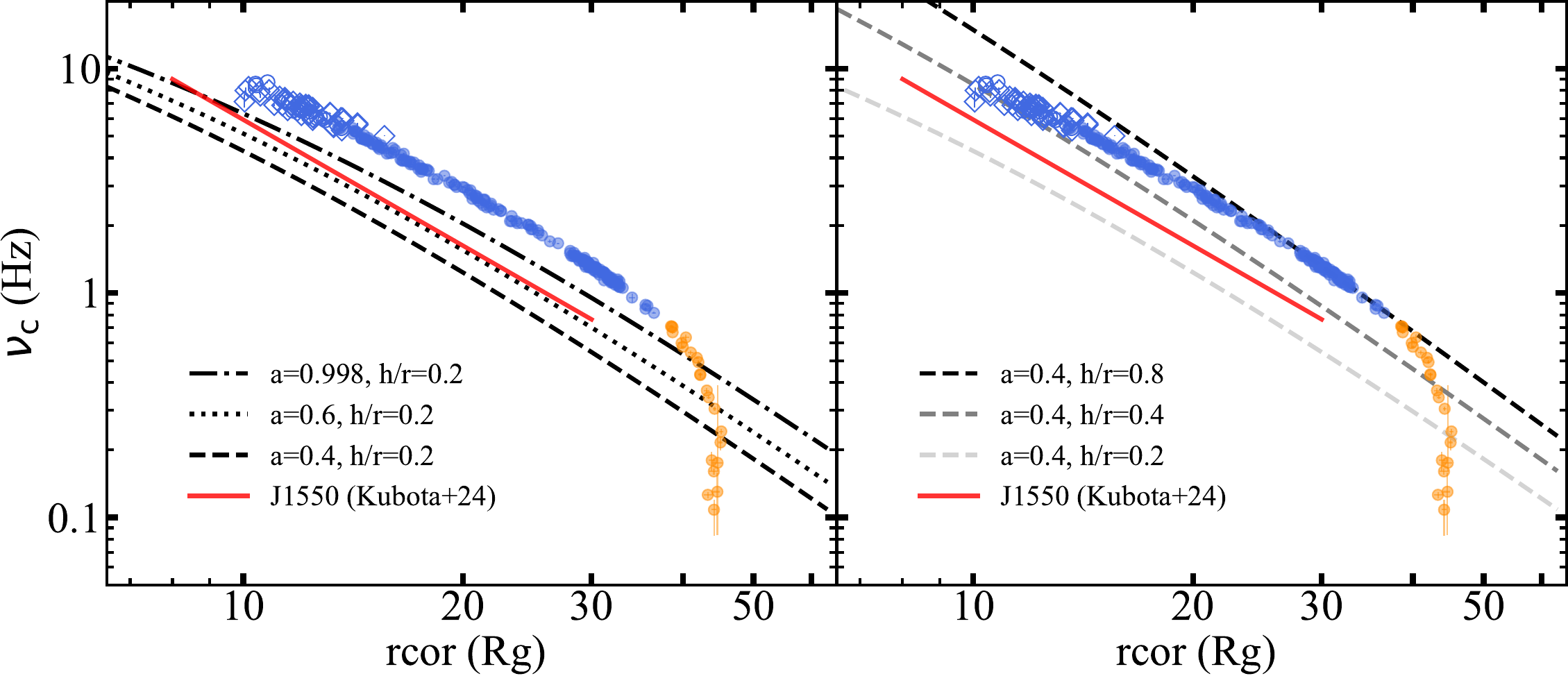} \\
\caption{The $\nu_{\rm c}$–$r_{cor}$ relation of Swift~J1727.8--1613. Orange data points correspond to the HS, while blue points represent the remainder. The red solid line represents the $\nu_{\rm c}$–$r_{cor} $ relation, excluding HS data, for XTE J1550--564 in \citet{Kubota2024}. Left panel: Dashed, dotted, and dash-dotted lines show the predicted $\nu_{\rm c}$–$r_{cor}$ relation from the LT model, with a fixed $h/r = 0.2$ and varying spin parameters $a_{*} = 0.4$, $0.6$, and $0.998$, respectively. Right panel: Similar to the left panel, but with a fixed spin parameter $a_{*} = 0.4$ and varying $h/r = 0.2$, $0.4$, and $0.8$, represented by light to dark grey dashed lines.}
\label{fig:lt}
\end{figure*}

Thus the radius $r_{cor}$ is mostly constrained from the spectral shape alone, and the fits show that this decreases as the source spectrum softens, from 45\,$R_{\rm g}$ in the hard state to $\sim 9$\,$R_{\rm g}$ in the softest spectrum seen in these data. 
Fig.\ref{fig:lt} shows this radius plotted against the QPO frequency, $\nu_{\rm c}$. Despite the source’s complex spectral behavior — characterized by multiple flares in the light curve (see Fig.~\ref{fig:lc_hid}) — the observed $\nu_{\rm c}$-$r_{cor}$ relation is remarkably clear. We compare this with the predictions of the LT model below.  

\subsection{The $\nu_{c}-r_{\rm cor}$ relation and  LT precession}

\label{sec:disc_LT}

In the LT model of \cite{Ingram2009}, the QPO originates from vertical precession of the entire Comptonising flow, from $r_o = r_{cor}$ to the inner radius of the hot flow $r_{\rm i}$.  The LT precession frequency is a strong function of radius, so unconnected rings at different radii precess at very different frequencies, so there is no coherent signal. However, in a hot accretion flow, the misalignment torques are communicated across the flow by bending waves, allowing the entire flow to precess globally on the average (mass weighted) precession frequency. \cite{Fragile2007} gives an approximate expression for this as:
\begin{equation}
\nu_{\rm c}\approx \frac{5-2\zeta}{\pi(1+2\zeta)}\frac{a_{*}[1-(r_{i}/r_{cor})^{(1/2+\zeta)}]}{r_{cor}^{5/2-\zeta} r_{i}^{1/2+\zeta}[1-(r_{i}/r_{cor})^{5/2-\zeta}]}\frac{c}{R_g}
\end{equation}
where $\zeta $ is the power law exponent of the surface density profile, assuming a power-law form $\Sigma(r)=\Sigma_0 r^\zeta$. Simulations give $\zeta\approx 0$, simplifying this relation. This gives
\begin{equation}
\nu_{\rm c}\approx \frac{2.9}{\pi}\frac{a_{*}^{0.8}(h/r)^{0.4} [1-(r_{\rm i}/r_{cor})^{1/2}]}{r_{cor}^{5/2} [1-(r_{\rm i}/r_{cor})^{5/2}]}\frac{c}{R_g}
\end{equation}

Importantly, in the LT model, the inner radius is set by the misalignment torques rather than by the ISCO, so $r_{\rm i}$ is not equal to $r_{\rm in}$ derived from the spectra but rather is set by where the flow is truncated by the bending wave torques at $r_i\sim 3 (h/r)^{-0.8} a_*^{0.4}$ \citep{Ingram2009}. We calculate the $r_{\rm i}$ value self consistently for each of the model spin and $h/r$ shown in Fig.\ref{fig:lt}. 

Fig~\ref{fig:lt}a shows the results compared with the LT predictions for $(h/r)=0.2$ for a variety of spins assuming a mass of $9M_\odot$. The observed $\nu_{\rm c}$-$r_{cor}$ relation slope matches extremely well with the LT predictions for a hot flow with constant $h/r$.

However, unlike the results for XTE~J1550--564, indicated by the additional red solid line, there is an offset by a factor $\sim 2$ from the predicted $\nu_{\rm c}$ values for a given spectral measure of $r_{cor}$. This indicates that either the system parameters (spin, mass, distance, inclination) used for Swift J1727.8--1613 are wrong, or that there is some real difference in source geometry. We discuss the effect of each system parameter below. 

Spin clearly impacts the LT prediction but only fairly weakly (see also \citealt{Ingram2009}). 
Fig~\ref{fig:lt}a shows a range of spins including maximal and still the data are offset above the predictions, so spin is not a likely origin for this offset. 

Instead, mass gives a linear offset, as the QPO frequency scales as $c/R_g$ so the factor 2 difference could be produced from a mass of $4.5M_\odot$, but this 
seems unlikely to be the solution not 
just because this is at the lower end of the mass function for a black hole, but because it seems fine tuned that $\nu_{\rm c}$ does not then extend to higher values than the $\sim 6$~Hz seen in type C QPOs in XTE J1550--564. 

A real change in source geometry such as $h/r$ of the hot flow shifts the correlation as $(h/r)^{0.4}$. 
Thus, to get the overall offset of a factor 2 change in $\nu_{\rm c}$ requires a factor $\sim$5.6 change in $h/r$, i.e., $h/r\sim 1$ rather than $0.2$. However, this seems unlikely: the hot flow needs to be non-spherical in order for the precession to modulate the X-ray flux, and in order to produce the fairly high levels of X-ray polarization seen in IXPE during the outburst \cite{Ingram2024}.

The only system parameter left is distance, and this impacts not on $\nu_{\rm c}$ but on $r_{cor}$. A smaller distance means the source is less luminous, so the disc temperature at a given radius 
is lower. We find that we need a source distance of $\sim 2.7$~kpc for the same black hole mass, spin and inclination.  This is within the range of current distance estimates 
to this source,  especially if the inclination is also allowed to be higher, so seems a likely solution.

The final alternative, where the LT model is not correct, also seems fine tuned given how well the slope of the $\nu_{\rm c}$-$r_{cor}$ matches the LT prediction. 

\subsection{The highest $\nu_c$: HIMS3}
\label{sec:disc_LT}

Towards the end of the observation, at the highest QPO frequencies, the fits abruptly become substantially worse, but the $r_{cor}$ derived from the model with fixed $r_{\rm in}$ still lies on the same relation. Appendix A shows that if we allow $r_{\rm in}$ to vary then these highest frequency QPOs instead lie on a separate branch. 
We prefer the smooth evolution shown here, as it seems more physical but at the 
very least this indicates that the approximations 
in our model and/or 
in the LT predictions break down when the precessing region is very small. This could be due to a variety of effects: there could be substantial additions torques between the disc and hot flow, which affect the predicted QPO frequency \citep{Bollimpalli2024}, or perhaps the torques may become so strong that the inner ring can break and/or flip over the black hole.
\cite{Liska2021} shows simulations where frame dragging effects 
induced by the black hole spin can overpower the viscous torques, causing the inner disc to undergo rapid differential precession and even break into distinct sub-discs. In cases where the disc is both thin and highly inclined, as appears to be the case for Swift J1727.8--1613, the inner regions can become misaligned with the black hole equatorial plane, leading to sudden changes in the inner disc radius and overall accretion structure.

Whatever the cause, this
highlights a fundamental discrepancy 
in our modeling. The inner edge of the hot flow in the LT model is set by the bending waves, with $r_{bw}\sim 8.2\,R_{\rm g}$ assuming the same parameters as in XTE J1550--564, i.e. $h/r\sim 0.2$, $a_*\sim 0.5$, while
the inner edge of the hot flow in the spectral model is set by spectral fitting to $r_{\rm in}=4.5$. The original simulations for the LT model showed that the hot flow truncates at $r_{bw}$, plunging into the black hole with little additional energy extraction below this point. \cite{Fragile2007} suggested explicitly that they expect that misaligned hot flows should be less radiatively efficient than aligned hot flows. 
Yet this is clearly not the case. The (almost) soft state spectrum here requires $r_{\rm in}\sim 4.5$ and this is disc dominated, so it cannot precess globally ($\alpha >h/r$), so it is not truncated by the bending waves. 

We speculate that the complex behaviour of the magnetic viscosity does allow the gravitational potential energy from $r_{bw}$-$r_{\rm in}$ to be emitted, but that this very inner part of the flow cannot precess. The type C QPO is then limited to $r_{cor}>r_{bw}$, so it stops when the disc reaches $r_{bw}$
but this is not when the disc reaches the ISCO. This could give a reason why the type C QPO ends at $r_{\ i}\sim 8\,R_{\rm g}$ ($r_{bw}$) 
rather than at the ISCO radius of 4.5\,$R_{\rm g}$. We urge new simulations to explore the flow behaviour in this regime.

\subsection{LT relation: hard state}
\label{sec:dis_hs}

In the early hard state, the data deviate from the LT prediction line, as shown in Fig~\ref{fig:lt}. While the source exhibits a low QPO centroid frequency, the $r_{cor}$ derived from SS{\sc sed} model fitting is too small to align with the LT prediction. As mentioned above, during the early hard state, the data deviate from the LT prediction line, as shown in Fig.~\ref{fig:lt}. Although the source displays a low QPO centroid frequency, the coronal radius ($r_{cor}$) derived from the SS{\sc sed} model fitting is too small to be consistent with the LT prediction. As noted in \cite{Kubota2024}, the SS{\sc sed} model is not well-suited for describing spectra in the HS. This is because the SS{\sc sed} model assumes a sandwich-type disk-corona geometry, in which a large number of seed photons are available. Consequently, it struggles to reproduce the observed hard spectra with $\Gamma \textless 2$. This implies that the assumption of a passive disk extending down to the ISCO no longer holds in the hard state, as long suggested in the truncated disc models for this state. In this geometry the 
seed photons could originate from the outer truncated disk instead of the passive disk assumed in the SS{\sc sed} framework. 
This remains a subject for future work.

\subsection{Jet power}

The SS{\sc sed} model is based on the assumption of the efficient radiation of the accretion power (half of the gravitational potential, with the remainder kept as kinetic energy of the inner disc). It is able to well describe most of the data seen here (apart from HIMS3) - with a fixed inner flow radius. Thus the majority of the data are consistent with this assumption, which is somewhat unexpected as there is a radio jet across all these observations
(see \citealt{zdziarski_2025} for days 0--20, and \citealt{Ingram2024} for the remainder), or \citet{Hughes2025} for the entire outburst. 

Most of this radio emission is from the steady compact jet, with radio flux which  rises from days 0--5 (hard state), then is relatively constant during the HIMS, dips by a factor of 3 during the X-ray peak in the flare state on day 27, then recovers but from day 29 onwards drops more or less monotonically by a factor of 10 through HIMS2-3, reaching a minimum at the soft state on day 43 \citep{Hughes2025}.

Thus there is a steady compact jet throughout these observations (as well as discrete ejections during the flare on day 27: \citealt{Wood2025}). This steady jet is an order of magnitude lower in HIMS3 than in HIMS1. This may be the origin of the jump in $r_{\rm in}$, as less jet power losses means more available power to radiate. However, the true soft state is consistent with $r_{\rm in}=4.5$, making it unlikely that the change 
in spectrum seen during HIMS3 is due to a change in jet power.

\section{Summary}
\label{sec:sum}

In this work, we analyzed the spectral and timing properties of the newly discovered black hole candidate Swift J1727.8--1613 using broadband (2--150 keV) observations from Insight-HXMT during its 2023 outburst. To investigate the evolution of the accretion geometry, particularly the truncation radius of the accretion disc, $r_{cor}$, and to test the applicability of the LT precession model in explaining QPOs. 

Using the energy-conserving SS{\sc sed} model, we performed a detailed spectral analysis, revealing two Comptonization components during the hard component dominant state. We tracked the evolution of the coronal radius, $r_{cor}$, which decreased steadily from 45\,$R_{\rm g}$ to 9\,$R_{\rm g}$ alongside a reduction in $h/r$.  This evolution is consistent with the source transitioning from the hard state to the intermediate state. Despite the complex nature of accretion, including multiple flaring events, our results reveal a robust relation between $r_{cor}$ and $\nu_{\rm c}$, i.e., aligning well with the slope predicted by the LT precession model. 

However, we observe an offset between the model prediction and the data. This discrepancies may arise from uncertainties in system parameters (most likely distance).
We also see a separate branch at the highest QPO frequencies which may point to more complex behaviour as the comptonisation region approaches the innermost radius. Some part of this is predicted in the LT models, where the hot flow is truncated at a bending wave radius which is somewhat larger than the innermost stable circular orbit. Alternatively, it may indicate more exotic behaviour 
such as disc tearing and/or breaking in the 
innermost accretion regions.

We show that the data can be mostly well fit by models where there is a constant radiative efficiency in the X-ray emitting flow. This is somewhat unexpected as the steady compact jet declines by a factor 10 
in radio flux across this period. This means that the jet is unlikely to take substantial power from accretion: either the jets are intrinsically low power e.g. \citet{Zdziarski2024} or are powered instead by tapping black hole spin.

\section*{Acknowledgements}

We thank the anonymous referee for useful comments that helped us improve the paper. RM thanks H. Liu for her helpful discussion. RM acknowledges support from the Royal Society Newton Funds. CD acknowledges STFC through grant ST/T000244/1 and a Leverhulme Trust International Fellowship IF-2024-020. AK acknowledges JSPS through grand 24K07098.

\section*{Data Availability}

The data for \textit{Insight}-HXMT underlying this article is available format at \textit{Insight}-HXMT website (\url{http://archive.hxmt.cn/proposal}; data in compressed format).





\bibliographystyle{mnras}
\bibliography{ref} 




\appendix

\section{Extra Pictures}

\begin{figure*}
\centering
\includegraphics[width=0.9\textwidth]{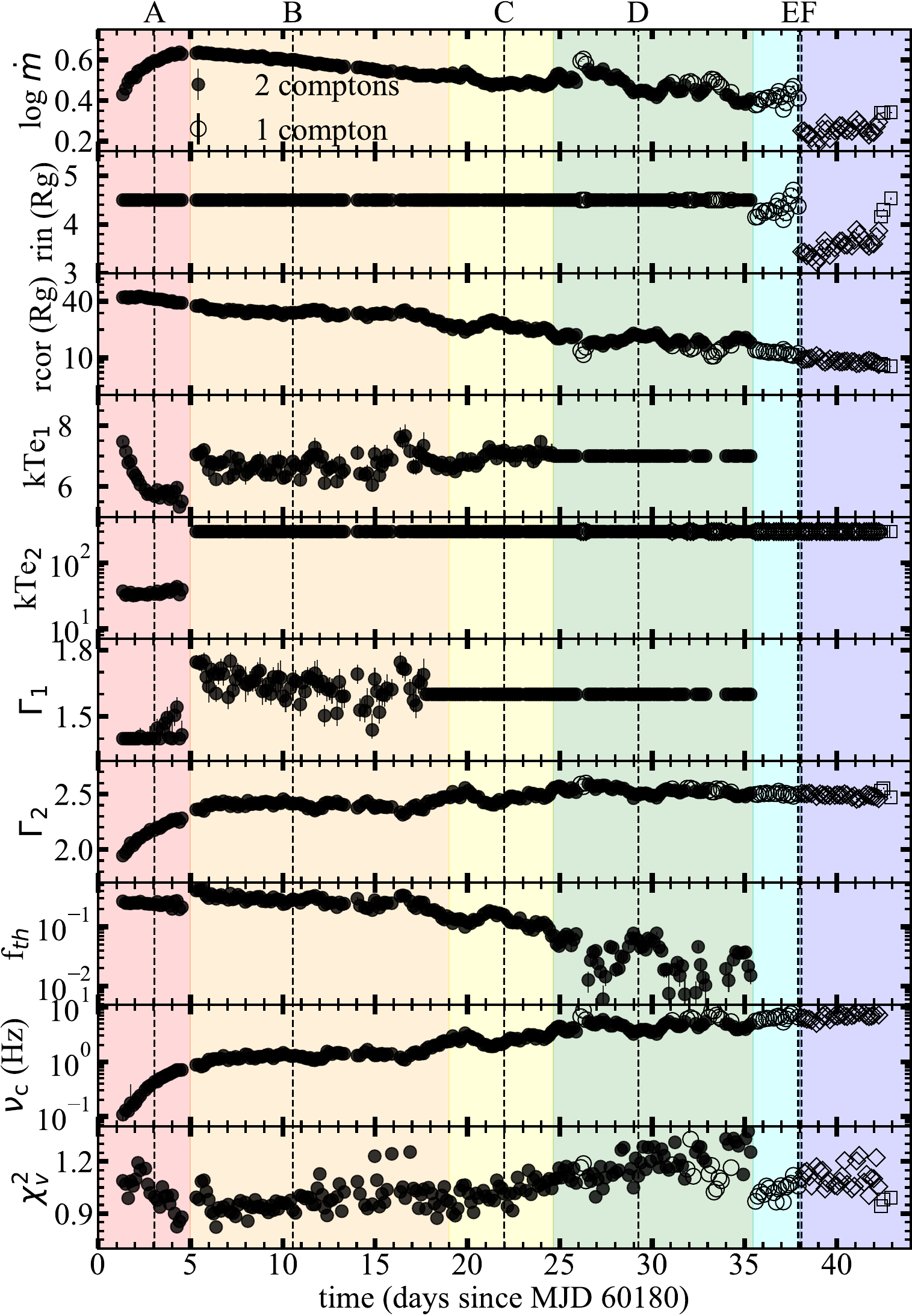} \\
\caption{Same as Fig.~\ref{fig:pars}, except that the HIMS2 and HIMS3 data are fitted with free $r_{\rm in}$. 
\label{fig:pars_appendix}
}
\end{figure*}

\begin{figure*}
\centering
\includegraphics[width=\textwidth]{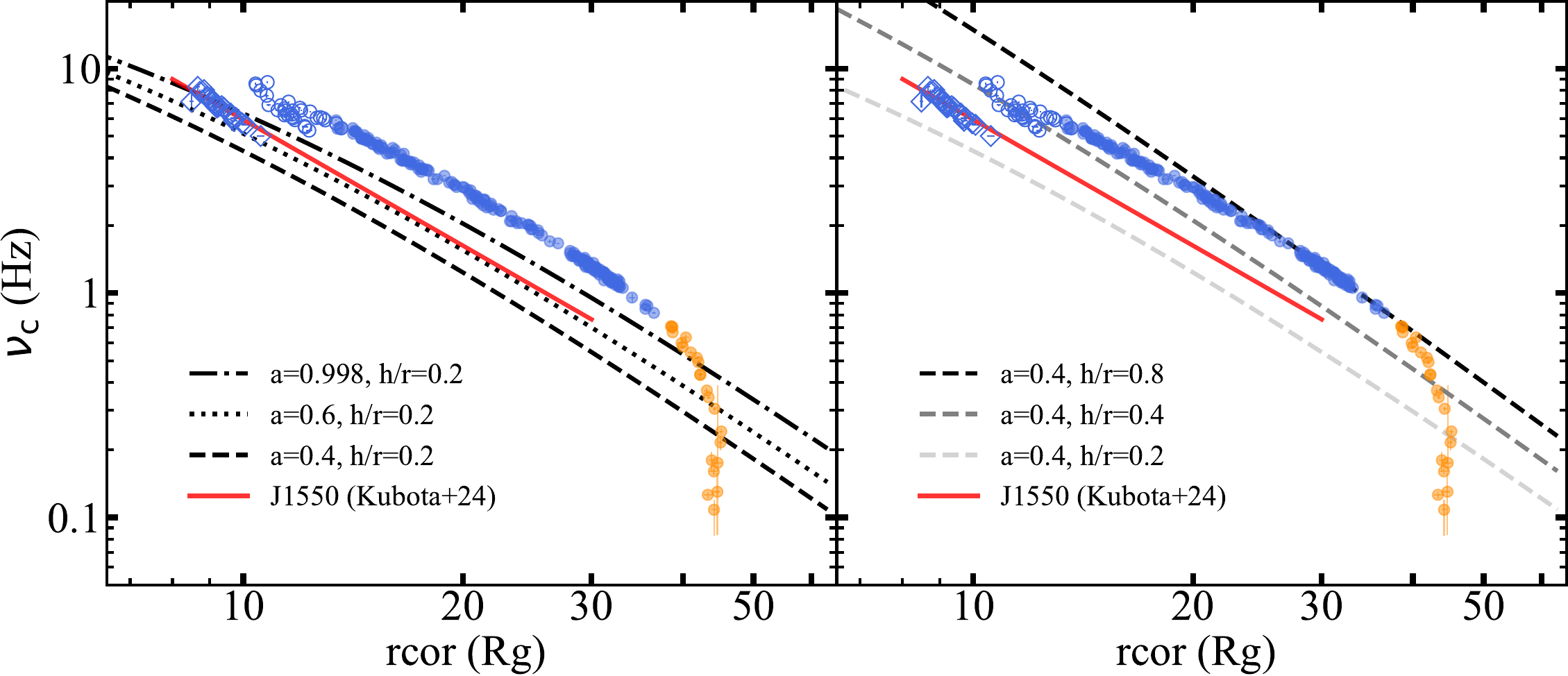} \\
\caption{Same as Fig.~\ref{fig:lt}, except that the HIMS2 and HIMS3 data are fitted with free $r_{\rm in}$. 
\label{fig:lt_appendix}
}
\end{figure*}

\section{harder Comptonization Component or Reflection Component?}
\label{sec:app_B}

We model the energy spectra by adopting a reflection component rather than a harder (thermal) Comptonization component. Following \citet{Kubota2024}, the reflected emission from the Comptonized continuum is modeled using the {\sc xspec} convolution model \texttt{xilconv}, which accounts for angle-dependent reflection from an ionized accretion disk. The \texttt{xilconv} model combines the ionized disk table model \texttt{xillver} of \citet{Garcia2013} with the Compton reflection code of \citet{Magdziarz1995}, and can be regarded as a modified version of the \texttt{rfxconv} model \citep{Kolehmainen2011}. This reflection component is further convolved with \texttt{kdblur}, which applies the general relativistic blurring expected for emission arising near a non-spinning black hole. The \texttt{xilconv} model also provides key physical parameters, including the reflection fraction ($ref\_refl$), iron abundance, inclination, and ionization parameter ($log\,\xi$). 

\begin{figure*}
\centering
\includegraphics[width=\textwidth]{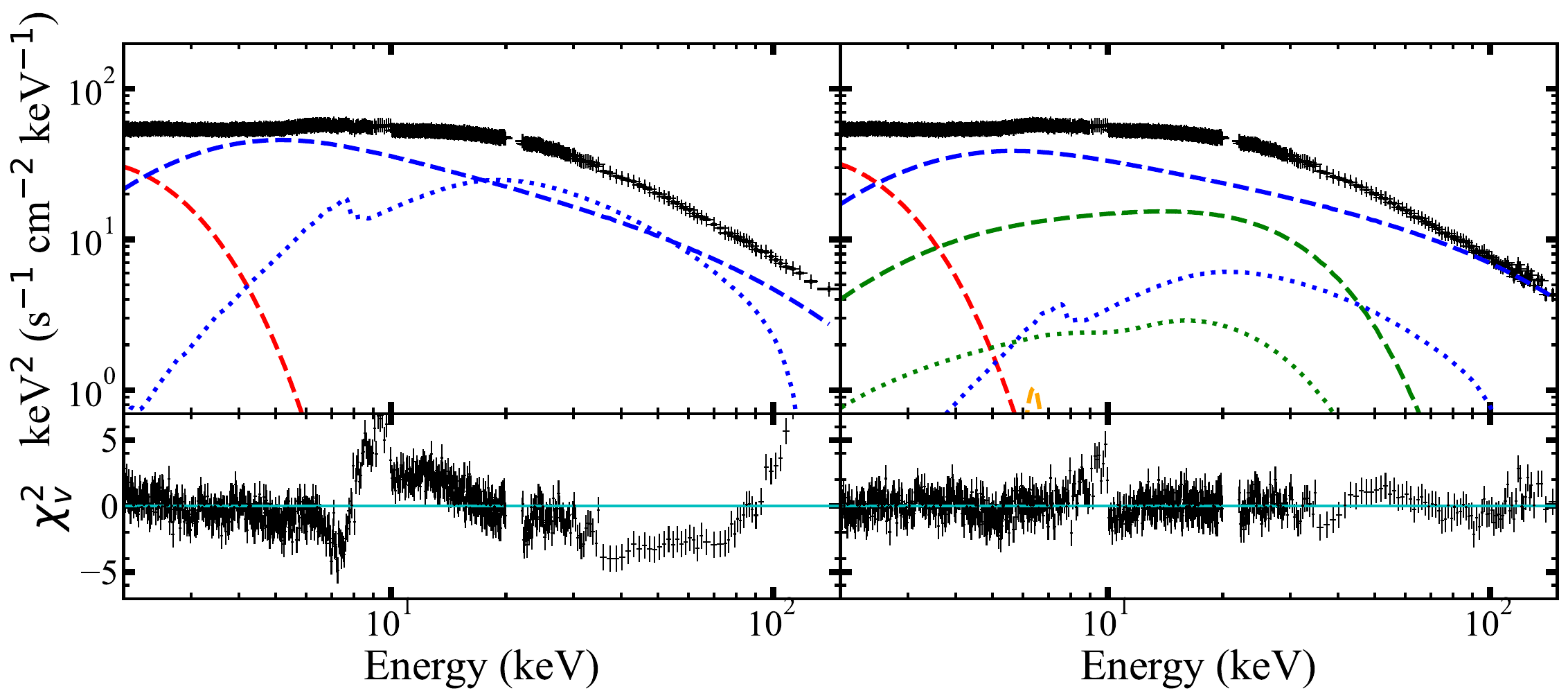} \\
\caption{Representative best-fitting energy spectra including reflection components, shown using the HIMS observation Obs. B (ObsID P061433800512) as an example. Left panel: the best-fitting model in {\sc xspec} is \texttt{tbabs * (SS{\sc sed}$_{\rm disk}$ + kdblur * xilconv * SS{\sc sed}$_{\rm 1\,nthcomp}$)}. Right panel: the best-fitting model in {\sc xspec} is \texttt{tbabs*(SS{\sc sed}$_{\rm disk}$ + kdblur * xilconv * SS{\sc sed}$_{\rm 2\,nthcomp}$ + gaussian)}. Line styles follow the same convention as in Fig.~\ref{fig:pha}, except that the blue and green dotted lines indicate the reflection components associated with the non-thermal and thermal Comptonization, respectively.}
\label{fig:fitpha_refl}
\end{figure*}

As shown in Fig.~\ref{fig:fitpha_refl}, fitting the energy spectra with a single Comptonization component and its reflection does not provide an adequate description, yielding $\chi^{2}/dof$=2985.36/1407. In contrast, a model with two Comptonization components and their reflections gives a significantly better fit, with $\chi^{2}/dof$=1187.82/1398. For clarity, we summarize the key parameters of the single-component model, the two-component model with reflections, and the two-component model without reflections in Tab.~\ref{tab:pars_refl}. Notably, for most of the spectral-fitting parameters, there is no significant difference between the cases with and without full reflection. For one of the key parameters under consideration, $r_{cor}$, the difference remains below 10\%.

We note that the heating and cooling rates underlying the balance in the \texttt{xillver} reflection models are currently being updated so there are some systematic uncertainties in these models which preclude more detailed studies of the reflection parameters \citep{Ding_2024}. The public release models also assume the photoionisation balance is only due to the coronal illumination from above which is not appropriate for most of the bright states studied here as they has substantial intrinsic disc flux \citep{Ding_2024}. We encourage future studies with more tailored reflection models.

\begin{table*}
\renewcommand{\arraystretch}{1.3}
\label{tab:pars_refl}
\begin{tabular}{lllll}
\hline
 Component & Parameter & (i) 1 nthcomp + ref  &  (ii) 2 nthcomp + ref & (iii) 2 nthcomp + gau  \\ \hline
{\tt SS{\sc sed}}  &  log $\dot{m}$ (${\dot Mc^{2}}/{L_{\rm Edd}}$)  & 0.49 &  0.56 &  0.60 \\ 
  &  $R_{cor}$ ($r_{\rm g}$)  & 24.8 & 28.1 & 30.2 \\ 
  &  $kT_{\rm e1}$ (keV)  & - &  6.89 &  6.80 \\
  &  $kT_{\rm e2}$ (keV)  &  [300] &  [300] &  [300] \\ 
  &  $\Gamma_1$  &  - &  1.79 &  1.67\\ 
  &  $\Gamma_2$  &  2.59 &  2.43 & 2.41 \\ 
  &  $f_{th}$  & - & 0.26 & 0.29 \\ \hline
{\tt kdblur, xilconv, gaussian} & $R_{\rm ref,in}$ (thermal nthComp) &  - & 1.51 & - \\ 
  &  $ref\_refl$ (thermal nthComp) & - & 0.30 & - \\ 
  &  $log\,\xi$ (thermal nthComp) & - &  3.29 &  - \\ 
  &  $R_{\rm ref,in}$ (non-thermal nthComp)  & 3.76 & 2.84 &  - \\
  &  $ref\_refl$ (non-thermal nthComp)  & 1.99 &  0.16 & - \\
  &  $log\,\xi$ (thermal nthComp)  & 1.00 &  1.01 & - \\ 
  &  $A_{\rm Fe}$ (solar)  &  0.5 &  0.5 &  - \\ 
  &  $E_c$ (keV)  &  - &  6.38 &  6.63 \\   
  &  $\sigma$ (keV)  & - & 0.29 & 0.56 \\ 
  &  $norm$ ($\times 10^{-2}$)  &  - &  0.8 & 5.1 \\ \hline 
  &  $\chi^{2}/dof$  &  2985/1407 & 1198/1398 & 1293/1405 \\ \hline 
\end{tabular}
\caption{Spectral parameters of the representative best-fitting spectra for Obs. B, obtained with different models: (i) a single Comptonization component with full reflection, (ii) two Comptonization components with full reflection, and (iii) two Comptonization components with a Gaussian line only.}
\end{table*}


\bsp	
\label{lastpage}
\end{document}